\documentclass[twoside, 10pt]{article}

\usepackage{lipsum} 

\usepackage[sc]{mathpazo} 
\usepackage[T1]{fontenc} 
\usepackage{microtype} 
\usepackage{graphicx}

\usepackage[top=1cm, bottom=0.5cm, left=1cm, right=1cm,columnsep=20pt]{geometry} 
\usepackage{multicol} 
\usepackage[hang, small,labelfont=bf,up,textfont=it,up]{caption} 
\usepackage{booktabs} 
\usepackage{float} 
\usepackage{epstopdf}
\usepackage{relsize}
\usepackage{multirow}
\usepackage{tabu}
\usepackage{array}
\usepackage{wasysym} 

\usepackage[labelfont=bf]{caption}

\usepackage[title,titletoc,toc]{appendix}

\usepackage{lettrine} 
\usepackage{paralist} 

\usepackage{abstract} 

\usepackage{titlesec} 
\renewcommand\thesection{\Roman{section}} 
\renewcommand\thesubsection{\Roman{subsection}} 
\titleformat{\section}[block]{\large\scshape\centering}{\thesection.}{1em}{} 
\titleformat{\subsection}[block]{\large}{\thesubsection.}{1em}{} 

\usepackage{wrapfig}

\title{\vspace{-15mm}\fontsize{19pt}{10pt}\selectfont\textbf{Machine Learning for Better Models for Predicting Bond Prices
}} 
\vspace{-0.3cm}
\author{
\large
\textsc{Swetava Ganguli, Jared Dunnmon}\\
{\{swetava, jdunnmon\}@cs.stanford.edu}
\vspace{-0.5cm}
}\vspace{-2cm}
\date{}

\begin{document}

\maketitle 


\vspace{-0.5cm}
\begin{abstract}
\vspace{-0.3cm}
\noindent Bond prices are a reflection of extremely complex market interactions and policies, making prediction of future prices difficult. This task becomes even more challenging due to the dearth of relevant information, and accuracy is not the only consideration--in trading situations, time is of the essence.  Thus, machine learning in the context of bond price predictions should be both fast and accurate. In this course project, we use a dataset describing the previous 10 trades of a large number of bonds among other relevant descriptive metrics to predict future bond prices. Each of 762,678 bonds in the dataset is described by a total of 61 attributes, including a ground truth trade price. \textbf{We evaluate the performance of various supervised learning algorithms for regression followed by ensemble methods, with feature and model selection considerations being treated in detail.  We further evaluate all methods on both accuracy and speed.  Finally, we propose a novel hybrid time-series aided machine learning method that could be applied to such datasets in future work}. 

\end{abstract}


\begin{multicols}{2} 

\section{Introduction}
\textbf{Key Problem:}\\
\\
Bond markets are generally characterized by a substantial dearth of trading information with respect to the amount of information available to equity traders. While equity traders can access stock bids, offers, and trades within 15 minutes of these activities, analogous information on bonds is only available to those who engage a fee-for-data contractor, and even then only in relatively small subsets compared to the overall volume of bond trades.  The asymmetry in required versus available information leads to the current state wherein many bond prices are in fact days old and do not accurately represent recent market developments \cite{Benchmark}. \\
\\
\textbf{Our Goal:}\\
\\
The goal of this project is to use the techniques and algorithms of machine learning and a set of data describing trade histories, intermediate calculations, and historical prices made available (on Kaggle) by Benchmark Solutions, a bond trading firm, in order to more accurately predict up-to-date bond prices using data that would be viable to obtain at a particular moment in time  \cite{Benchmark}. The high volume of data characteristic of this problem is common in such financial modeling endeavors, and hinders the formation of fully descriptive a priori theoretical models.  In this report, we develop strategies to effectively utilize the data provided for bond price prediction via thorough investigation of the space of available machine learning models and combination with methods from time-series analysis.\footnote{Computing time on Stanford Corn, Barley and Rye clusters is gratefully acknowledged}\\
\\
\textbf{Strategy and Methods:}
\begin{description}
	\item[Feature Selection:] An important aspect of this task is creating class-balanced training and test data sets while identifying appropriate metrics for assessment of prediction success. Critical features are analyzed and extracted using low order modeling techniques like Principal Component Analysis (PCA) and correlation analysis. 
	\item[Supervised Learning Methods:] We first investigate computationally inexpensive techniques such as Generalized Linear Models (GLMs) and regression trees. We also assess the viability of methods like Principal Component  Regression (PCR) and Support Vector Regression (SVR). 
	\item[Ensemble Methods:] Since we have a regression problem at hand, regression trees are combined as weak learners in ensemble methods like Bagging, LS-Boosting and Random Forests to reduce overfitting and to potentially take advantage of the large size of the dataset. 
	\item[Hybrid Time-Series Methods] Because each bond includes historical data on five different quantities for the last ten trading periods, we investigate the possibility of feature space augmentation or reduction using Time-Series (TS) analysis.  Ideally, predictions from TS methods would either provide new features with additional explanatory power or enable reduction of the feature set size while retaining explanatory power. 
	\item[Neural Networks:] We experiment with applying neural networks to this problem, as they are known to fit even highly nonlinear data well given sufficient neurons. 
\end{description}

\vspace{-0.3cm}
\section{Exploratory Data Analysis}

The data used for this project contains 61 attributes observed for each of 762,678 bonds: 3 Nominal, 12 Discrete Ordinal, 1 Observation Weight and 45 Continuous (Ratio) Attributes, including a ground truth trade price.  To predict the bond price (often called the "trade price"), the data delineates a unique ID of the bond (nominal discrete attribute), a categorical ID of the bond (nominal discrete attribute), a weight/importance of each bond (continuous ratio attribute), the bond coupon (continuous ratio attribute), years to maturity (continuous ratio attribute), whether the bond is callable or not (nominal discrete binary variable), seconds after the trade occurred that it was reported (continuous ratio attribute), notional amount of the trade (quantitative discrete attribute), the type of trade that occurred (2 = customer sell, 3 = customer buy, 4 = trade between dealers), and a fair price estimate based on implied hazard and funding curves of the bond issuer (continuous ratio attribute).  This last attribute is referred to from hence forth as the "curve-based price." In addition, the dataset also has information about the last 10 trades that occurred on each bond considered, including the time difference between a trade and the previous trade (continuous ratio attribute), the trade price (continuous ratio attribute), the notional trade amount (continuous ratio attribute), the trade type (binary discrete nominal attribute), and the curve-based price (continuous ratio attribute). \\
\\
\textbf{Correlated Attributes:}\\
\\
We observe from the correlation matrices that attributes \textit{Price of the Last Trade} and \textit{Curve-Based Price of the Last Trade} are strongly correlated at all time points. This is intuitively expected. Thus, this information can be used to inform dimensionality reduction. 
The fact that the remainder of the variables are minimally correlated implies that each of those attributes should supply new information for our prediction.  A similar conclusion can be observed when autocorrelations are computed for these different time series. 
Specifically, the mean autocorrelations for each variable are very low ($\rho<0.3$) beyond the first lagged period, indicating that each variable contributes unique information at every time period.\\
\\
\textbf{Treatment of Categorical Attributes:}\\
\\
Empirical PDFs of the nominal attributes have been analyzed. From the PDF of the attribute that denotes whether the bond is callable, we see that 89 \% of the bonds are not callable whereas 11 \% of the bonds are callable. From the empirical PDF of the attribute \textit{trade type of the current trade}, it is seen that the current trade has 20 \% of the type 2 trade, 36 \% of the type 3 trade and 43 \% of the type 4 trade. Thus, there is a relatively uniform sampling of the three trade types. While preparing the cross-validation datasets, this will be taken into account such that they are class-balanced. Furthermore, in the ensemble methods used (which are regression tree-based since we are predicting a a continuous output), these categorical variables are handled appropriately in cases where they are nominal or ordinal.   

\section{Cross-Validation, Feature Selection and Model Evaluation Metric}
\begin{figure}[H]
\centering
\includegraphics[width=0.4\textwidth]{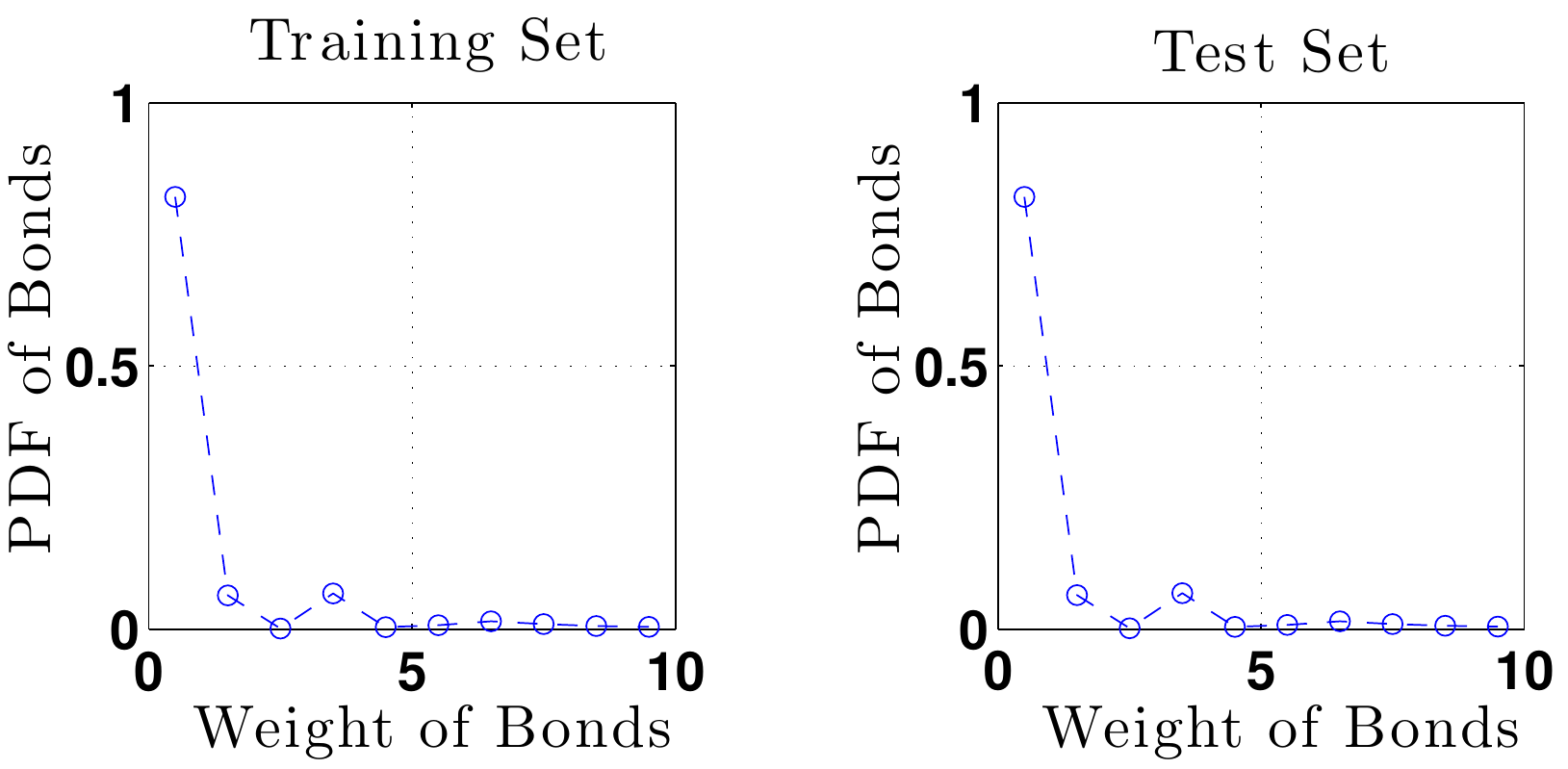}
\caption{\normalsize{Demonstration of weight-balanced training and test datasets.}}
\label{Figure2}
\end{figure}

\textbf{Cross-Validation Strategy:}\\
\\
The key problems encountered in the process of feature selection and in creating training and test data sets are:
\begin{enumerate}
	\item The time series length is either on the borderline or below the minimum number of points required for a statistically consistent time-series prediction
	\item Categorical attributes have non-uniform distributions
	\item The amount of data characterizing the various categories of bond importance is distinctly non-uniform
\end{enumerate}
Points 1 and 2 are direct manifestations of the dearth of data for bond price prediction. It is also important to correctly predict the highly weighted bonds well since they are usually of higher priority in a portfolio. Due to the problems mentioned above, it is difficult to conduct typical k-fold cross-validation wherein the training sets would be class-balanced. Instead, in order to utilize all the data given, we utilize a 70-30 hold-out cross-validation.  We therefore create weight balanced training and test sets using the following algorithm.\\
\\
\textbf{Algorithm for Cross-Validation:}
\begin{description}
	\item[Step I] Randomly create 5 instances of weight balanced training and test sets. 
	\item[Step II] Run Machine Learning Algorithm on each of these 5 training and test sets.
	\item[Step III] Report the appropriate metric (discussed below) from each of the 5 independent runs.
	\item[Step IV] The final value of the evaluation metric is the average of these 5 values.
\end{description}
To demonstrate that our sets are indeed weight balanced, we plot the PDF of the bond weights for one instance of the training and test sets in Figure \ref{Figure2}.
\noindent \textbf{Model Evaluation Metric:}\\
Given that errors in bond pricing are equally detrimental in upward and downward directions \cite{Hull2011}, we evaluate our predictions based on the simple weighted $L_1$-norm of the difference between the actual price and our predictions per sample (i.e. per bond). Thus, the \textbf{model evaluation metric} that we choose is the \textbf{Weighted Error in Prediction per Sample (WEPS)} which is defined as:
\begin{equation}
\mathlarger{WEPS = \frac{\sum_{i = 1}^{m} w_i(\big|y_{true} - y_{predict}\big|)}{\sum_{i = 1}^{m} w_i}}
\end{equation}~\\
Note that all prediction errors are calculated using our cross-validation algorithm.\\
\\
\noindent \textbf{Statistical Significance:}\\
Let $e_1$ and $e_2$ be the errors obtained from two different models $M_1$ and $M_2$. Since the number of records in the training datasets and test datasets for all models is the same, say $n$, we can write the observed difference in the error as: $d = e_1 - e_2$. The variance of $d$ can be computed as
\begin{equation}
\sigma_d^2 \approx \hat{\sigma}_d^2 = \frac{1}{n} (e_1(1 - e_1) + e_2(1 - e_2)) \sim \mathcal{O}(10^{-6}) \vspace{-0cm}
\end{equation} 
The 95\% confidence interval in our case is then given by ($d_t$ = True Difference)
\begin{equation}
d_t = d\,\,\pm\,\,1.96\hat{\sigma}_d
\end{equation}~\\
Importantly, this implies that any improvements in the WEPS metric out to the fifth decimal place are indeed statistically significant.\\
\\
\textbf{Feature Generation and Selection:}\\
\\
Feature selection and generation is handled as follows:
\begin{description}
	\item[Correlation Analysis:] No attributes supplied are strongly correlated. Mild correlations exist in only 2 sets of 		attributes. Thus, this method is not particularly informative.
	\item[PCA in Supervised Learning:] PCA is run on the full dataset with the goal of determining if there exists a reduced feature set that retains the majority of the explanatory power of the full feature set.  
	\item[Scoring Function for Ensemble Methods:] Random Forests (RF) are used for feature ranking.  RF will select features randomly with replacement and group every subset in a separate subspace  (called  the random subspace). We use a scoring function with the following methodology. If feature $X_2$ appears in 25\% of 	the trees, then score it. Otherwise, we do not consider ranking the feature because we do not have sufficient 			information about its performance. We then assign the performance score of every tree in which $X_2$ appears to $X_2$ and average the score. Our search method is recursive: For example, if we drop the worst 20\% in the first round, we do so in all following rounds until the desired number of features is attained. The 20\% parameter has been determined via numerical experiment.
\end{description}

\section{Models from Supervised Learning}
\begin{figure}[H]
\centering
\includegraphics[width=0.35\textwidth]{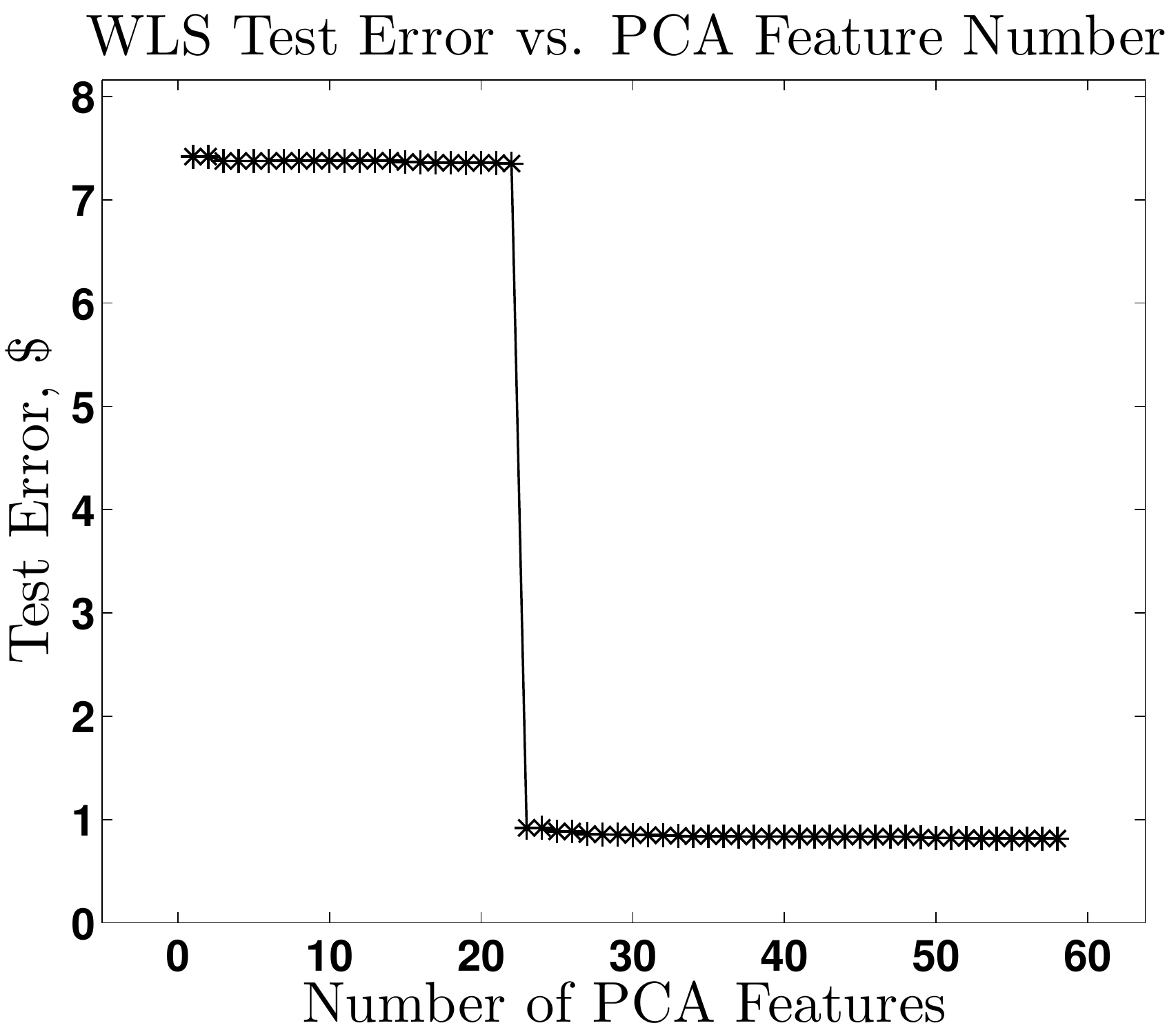}
\caption{\normalsize{WLS WEPS versus PCA Feature Number.}}
\label{PCARedFig}
\end{figure}

We now proceed to explaining implementation and performance of the various models utilized here.  All algorithms were implemented in Matlab for ease of workflow, and all results referenced in the text can be found in Figure \ref{resultstable}. \\
\\
\textbf{Generalized Linear Models:}\\
\\
Several models from supervised learning were investigated.  First, an unweighted generalized linear model was implemented using two different link functions and the full feature set in order to investigate the underlying distribution of the data.  While financial data often has an underlying normal variation, it is important to ensure that this assumption is valid before proceeding.  We report the results of Ordinary Least Squares (OLS) regression using link functions for the normal and gamma distributions.  Evaluating the training and test errors for these different cases illustrates that normal variation appears to best characterize the data.  To improve on these results, Weighted Least Squares (WLS) was performed using the evaluation weights to appropriately govern which points are treated with highest importance in the regression.  WLS gives noticeable 3.1 \cent $ $ improvement over OLS (in the context of errors on the order of \$1). \\
\\
\textbf{Principal Component Regression:}\\
\\
We next proceed to implementing PCR using the reduced feature set.  We reduce the size of the feature set to 23 features using this procedure, as the PCA routine reports that all principal components higher than 23 are nearly linearly dependent.  In order to make a prediction based on PCA, we extract the transform utilized in the PCA algorithm and apply this directly to the test data.  Once the data is transformed in this way, we can run GLM models as usual.  It was explicitly confirmed that transforming the regression coefficients back to the original covariate space gives the same predictions for OLS, validating the prediction procedure we use.  Interestingly enough, despite the fact that the first few principal components tend to explain variance in the input best, it is in fact the independent principal component in our data with the lowest eigenvalue (i.e. the last one in the reduced feature set) that provides the vast majority of the explanatory power with respect to the bond price.  This is illustrated explicitly in Figure \ref{PCARedFig}. Investigation of this feature's constituents reveal that it exclusively contains all of the historical curve and trade prices, implying that these variables have substantial impact on correct prediction.
In fact, the most explanatory PCA variables are so potent that while using 23-feature WLS gives an error of \$0.9191 in 12 seconds, a 3-feature WLS using only the three most explanatory PCA components gives a WEPS of only \$1.2637 in just 3 seconds.  This 75\% reduction in speed could certainly become important in dealing with massive datasets often encountered in this area of finance.\\
\\
\noindent \textbf{Support Vector Regression:}\\
\\
We briefly investigate the possibility of using Support Vector Regression (SVR) to predict bond prices, but found the model estimation process to be so time-intensive that it precluded effective parameter tuning.  Specifically, the LibSVM package did not report an SVR result in 5 days of computation time while the LibLinear package took 3 days just to estimate a single model.  It is possible that this has to do with the size of the memory cache allocated to storage of the support vectors.  Regardless, given these model estimation times, performing a parameter sweep over the critical SVR model parameters proved impractical given the time constraints of the project, and results for this method are therefore not reported.\\
\\
\noindent \textbf{Regression Trees:}\\
\\ 
Regression trees are known to overfit the data, thereby forming highly biased predictors. This behavior is observed in our analysis as well. The WEPS on the training set is extremely low (0.5\cent$\,$to$\,\,$1\cent) while the WEPS on the test set is \$2.5 to \$3. To mitigate overfitting, we experiment by (i) changing the number of data samples required at each node to make a decision, (ii) implementing different metrics for growing and pruning the trees (e.g. entropy, misclassification errors, etc.), (iii) varying the number of predictors randomly sampled at each node to make a decision, and (iv) controlling the depth of the trees. However, the test WEPS does not decrease in any of these cases.  A full list of tested conditions and representative results can be found in Figure \ref{resultstable}. 
\section{Models from Ensemble Methods}
We use Bagging of data samples, LS-Boosting (Sequence of Decision Trees), and Random Forests (Bagged Trees) as our ensemble regression methods with regression trees as the weak learner. Again, note that all prediction errors are calculated according to our cross-validation algorithm.\\
\\
\textbf{Random Forests with Regression Trees:}\\
\\ 
Random forests can be a useful method for feature selection as outlined above. However, they are computationally very expensive. Furthermore, they generally produce a WEPS of slightly above \$1, which is inferior to even simple models like GLMs.  Various methods have been experimented with to reduce the overfitting in each weak learner and to accelerate the convergence of the random forest algorithm, including: (i) changing the number of data samples required at each node to make a decision, (ii) different metrics for growing and pruning the trees (e.g. entropy, misclassification errors, etc.), and (iii) number of trees grown for the majority vote. However, none of these methods improves the performance of the random forest. The training and test errors along with the execution time shown in Figure \ref{resultstable} are characteristic of the experiments we conducted. \\
\\
\textbf{LS - Boost with RT as Weak Learner:}\\  
\\
We also experimented with the LS Boost algorithm with regression trees as the weak learner. Boosting is known to perform well due to exponential penalizations of observations proportional to the error in their prediction. $J$, the number of terminal nodes in trees, is the critical parameter that can be optimized for a given dataset. Hastie et al. \cite{Hastie} comment that typically $4 \leq J \leq 8$ works well for boosting and results are fairly insensitive to the choice of $J$ in this range. We therefore choose $J=6$. This algorithm performs better than random forests in that WEPS stagnates at around $80$\cent $\,$ and the computation time is 3 times lower. Characteristic values of WEPS on the training and test sets after performing cross-validation are shown in the summary table of Figure \ref{resultstable}.

\section{Models from Hybrid Time-Series Methods}
A relatively uncommon idea that was explored during this project was the possibility of using time series analysis to supplement the given feature set. Ideally, using a time-series method to make a prediction for the new price could allow for much of the historical data contained in the five time series for trade price, curve price, trade type, trade size, and time delay to be incorporated in a concise fashion. The goal of our work was to implement a time-series forecasting method that would allow us to create a set of time-series predictions that could be used to either augment our feature set or even replace all of the historical features in a concise fashion.

One issue that is immediately apparent with the dataset utilized here is that each bond contains time-series data for the last ten instances of the five variables mentioned above. Ten points is a very small number for normal time-series predictions, to the point that inbuilt Matlab packages for many standard time-series analyses cannot estimate model parameters with any degree of certainty from such a small dataset. Knowing this, we proceeded to investigate several potential options for time-series analysis in order to find even a very simplistic model that could at least output a reasonable prediction in the majority of cases with the goal of using it for feature-set augmentation as discussed above. For the sake of brevity, these complex time series models will be concisely described below, with provided references detailing full model specification. \\
\\
\textbf{Cointegration Models}\\
\\
One method of time-series prediction involves a procedure known as cointegration analysis. Briefly, a set of univariate time series can be considered cointegrated if some linear combination of these series and their lags is statistically stationary in time. A linear combination of time series (and lags) that is cointegrated is known as a cointegration relationship. If a set of cointegration relationships exists amongst a group of time series, these relationships can be used to forecast values for a subset of the time series at later points using historical data. The canonical statistical test for determining whether or not a set of time series data contains a cointegration relationship is the Engle-Granger test \cite{Shumway}.

The fundamental assumption of the Engle-Granger test is that a time-stationary linear combination of two time series, $y_t$ and $z_t$, is required for cointegration, such that,
\begin{equation}
y_t-\beta z_t = u_t, \nonumber \\
\end{equation}
with $u_t$ stationary. If $u_t$ were known \textit{a priori}, one could use an established statistical method such as the Dickey-Fuller test to evaluate stationarity \cite{Shumway}. However, in this case, we estimate $u_t$ using ordinary least squares and analyze the stationarity of the estimated series. A second iteration of this procedure is performed on the first differences of the each time series with the lagged residuals included. The combined output of the stationarity tests is presented as a single test statistic that can be used to evaluate the existence of a cointegration relationship.

We performed the Engle-Granger (E-G) test in Matlab on the time-series data from each of the nearly one million bonds in our dataset. Given that we only have ten points for each quantity for each bond, it is not unexpected that the majority of the E-G tests reported a p-value substantially above any reasonable significance boundary. In other words, this result means that it is generally not possible to reject the null hypothesis of there being no cointegration relationship amongst the different time series  variables describing each bond. This makes it difficult to specify a cointegration-based model such as a Vector Auto-Regressive (VAR) or Vector Error-Correction (VEC) model to forecast bond price based on cointegration relationships amongst previous time-series quantities.
Notably, however, there was a nontrivial proportion of the data (around 10 \%) for which the E-G test did allow for confident rejection of the null, implying the existence of at least one cointegration relationship. This result suggests that with access to additional historical data, it might well be possible to form a viable prediction for each bond that would allow for reduction of said historical data to a single time-series prediction that could then be input into the machine learning model discussed here \cite{Shumway}.
\\\\
\textbf{Auto-Regressive Moving Average (ARMA) Models}\\
\\
Another, slightly simpler method for approaching time-series forecasting is to predict future values of a variable based on its historical behavior. While there exist a wide variety of methods for accomplishing this task, ARMA models are quite common due to their simplicity and intuitiveness. In particular, the ARMA model incorporates two separate modes of capturing time-series behavior. The first can be summarized by considering that recent values of a time-series variable can be very good predictors of the present value. This is captured by the AR (Auto-Regressive) parameter in the model. The second line of thought, the Moving Average (MA) portion of the model, captures the fact that a large shock at a previous period would not only affect that period, but periods in the near future as well  \cite{Shumway}. The combination of these two models can be written as the following in terms of a time-series $y_t$, AR coefficients $\phi_i$, MA coefficients $\theta_j$, and deviations from the AR model $\epsilon_t$,
\begin{equation}
y_t = \sum_{i=1}^p\phi_iy_{t-i}+\epsilon_t+\sum_{j = 1}^q \theta_j\epsilon_{t-j},
\end{equation}
where $p$ and $q$ are the specification parameters of an ARMA($p,q$) model. Specification of $p$ and $q$ can be reliably achieved using the Box-Jenkins methodology, a well-documented procedure using easily computed autocorrelation functions\cite{Shumway}. The form of several visualized autocorrelation and partial correlation functions implies that an ARMA(1,1) model would potentially be appropriate for this data. Given the results of the cointegration analysis and the fact that the majority of observations allow statistically viable fitting to an ARMA(1,1) model, we pursue this approach in our time-series modeling.
In terms of results, the ARMA model was utilized in the following fashion. One of the data series provided, but not originally used in the regression is a categorical variable that is a relatively non-informative bond ID number. This variable identifies bonds that are of the same class (issuer, period, etc.). Because fitting an ARMA(1,1) model to each observation independently would take substantial computational resources and thus might not be the most helpful from a predictive standpoint, we used knowledge gained from our PCA analysis to define a method for efficiently integrating a time-series analysis into our models via this bond ID number. Specifically, our previous analyses indicated that historical trade and curve prices were responsible for most of the explanatory power of the model. We therefore aimed to use time-series estimates of the difference between the trade price and the curve price to extract additional information that would improve our results. Our algorithm is as follows:\\
\\
(i) Estimate an ARMA(1,1) model for 10 samples of each bond type in the training set using a variable defined as the difference in the trade and curve prices at each time point\\
(ii) Average ARMA parameters to create an average TS model for the difference in trade and curve price for that bond type\\
(iii) Forecast one period forward from the historical data, which gives a prediction of the difference between the trade and curve price for each bond type for the prediction period\\
(iv) Use this forecast variable as a new feature in GLM models\\
\\
The ultimate goal of this procedure was to integrate given data on the type of the bond in a manner more consistent with fundamental economic behavior as opposed to a simple categorical label. Theoretically, this series should have greater explanatory power than the bond identification number series alone because it incorporates time-series data. We illustrate this in practice by performing a reduced-feature set GLM\footnote{Referred to as "9-Feature WLS" in Figure \ref{resultstable}} using only data from the current time period, the curve and trade price from the first historical period, and either the bond identification number or the ARMA(1,1) variable. As shown in Figure \ref{resultstable}, including the bond identification number changes training and test error only by 0.04 \cent $\,$ and 0.01 \cent, respectively, for this reduced-feature case while inclusion of the ARMA(1,1) variable instead lowers training and test error by 2 \cent $\,$ and 2.3 \cent, respectively. Inclusion of the ARMA variable in the full WLS model similarly yields a respective reduction in training and test error by 0.5 \cent $\,$and 0.6 \cent. The fact that inclusion of the time-series variable enhances the performance of the GLM in both training and test errors suggests that this variable does adds new information to the model instead of simply causing overfitting. Importantly, if the time series model for each bond type is precomputed, it is a simple matter to use this feature to provide supplementary information about the bond price evolution in a simple GLM.
\section{Neural Networks}
Neural networks (NN) are very well-suited for function fitting problems. A neural network with enough neurons can fit any data with arbitrary accuracy. They are particularly well suited for addressing non-linear problems. We therefore experiment with two-layer (one hidden layer, one output layer) neural networks trained with the Levenberg-Marquardt optimization algorithm and simple back-propagation. The training and testing errors along with the execution time from this exercise are shown in Figure \ref{resultstable}. Two-Layer NNs perform quite well on our dataset, reducing test error to 73 \cent$\,$ in only 2 hours. WEPS reduction with network size beyond 20 neurons is quite gradual.
\begin{figure}[H]
\centering
\includegraphics[width=0.5\textwidth]{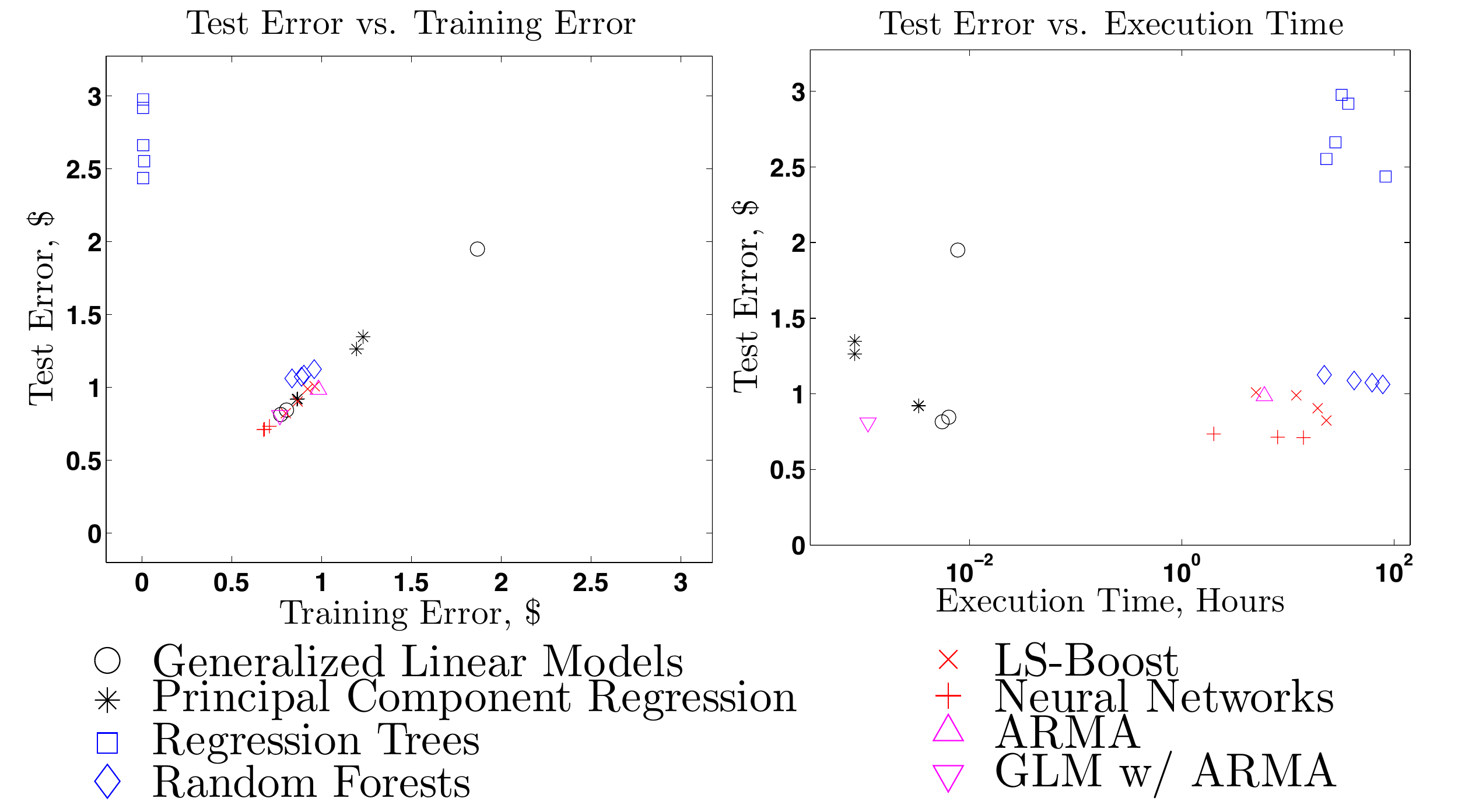}
\caption{\normalsize{Test Error versus Training Error and Training Time.}}
\label{Figure4}
\end{figure}
\section{Conclusions and Future Work}
At this point, we can make several definitive conclusions regarding the relative performance of the tested models in predicting bond price: (i) GLM models perform well with low computational cost (order of seconds), (ii) Feature set augmentation with TS models improves results, (iii) Ensemble methods do not substantially improve results, and require much more computational investment, (iv) Neural networks give very accurate results without overfitting in reasonable amounts of time (order of hours), (v) \textbf{NNs and GLMs give best results in terms of combined speed and accuracy}.\\
There exist several fruitful directions in which to take future work.  First, obtaining a dataset with longer time histories would allow for statistically significant specification of more detailed time-series models for improvement of feature augmentation.  Investigating the performance of different classes of time-series models as machine learning feature generation mechanisms would be useful.  Second, the success of neural networks on this dataset implies that investigating the application of multilayer networks and deep learning methods to this problem may yield better bond price predictions.  Finally, exploiting parallel implementation of these algorithms for model tuning would greatly enhance our ability to make the most accurate predictions possible.  All of these routes could yield improvements to our current results, and we intend to investigate several of these in the coming months.  
\section{Summary of Algorithm Performance}
The performances of the various methods are summarized below. The compute time is evaluated by running the code on one node on the Stanford Corn cluster.
\begin{figure}[H]
\centering
\includegraphics[width=0.5\textwidth]{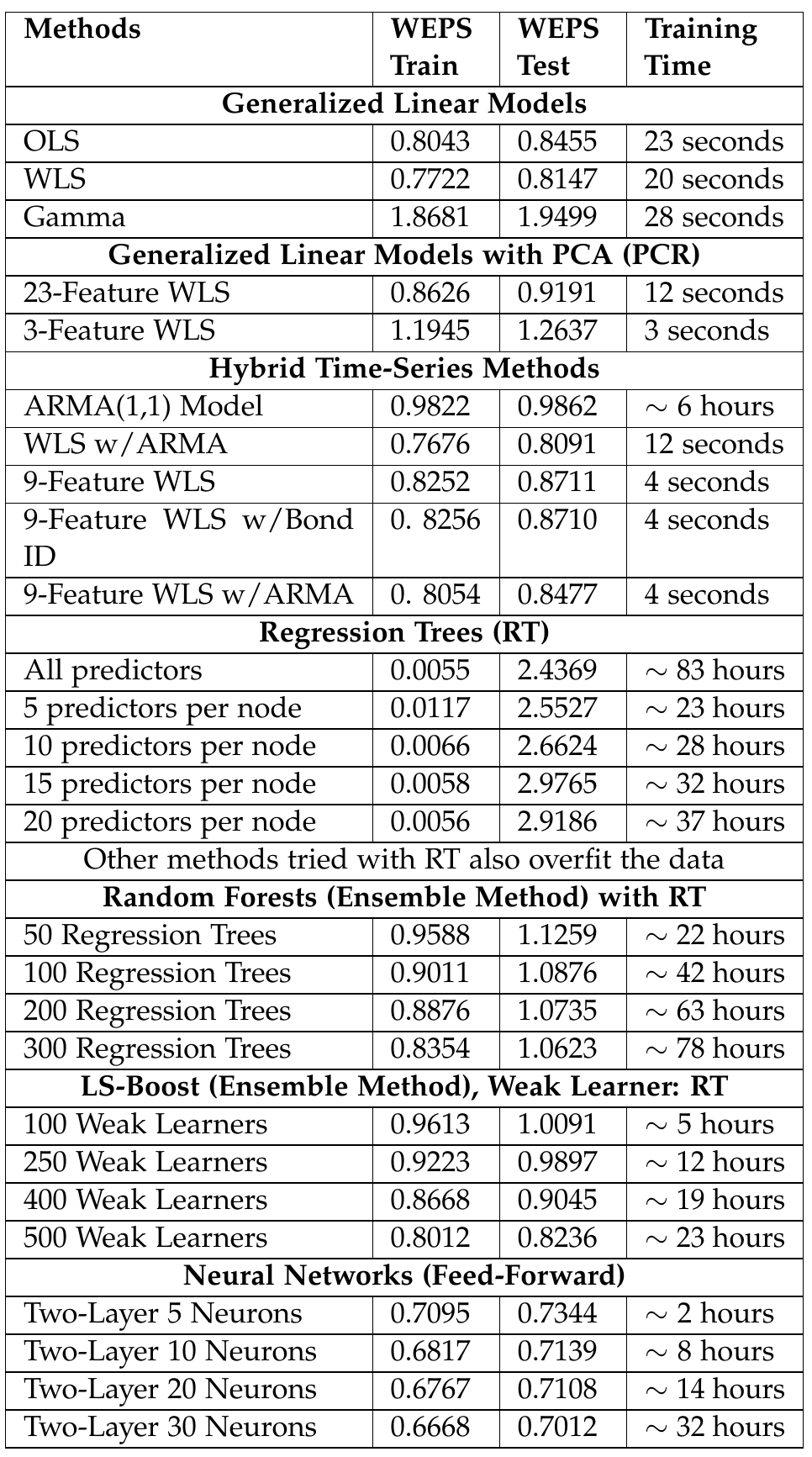}\vspace{-0.3cm}
\caption{\normalsize{Summary of Results.}}
\label{resultstable}
\end{figure}~\\
\vspace{-1.5cm}

\end{multicols}

\end{document}